\documentclass[twoside,12pt]{article}
\usepackage{epsf,epsfig,amssymb,amsmath,graphicx,float}
\usepackage{color}

\begin{document}
\renewcommand{\thefootnote}{\fnsymbol{footnote}}

\begin{titlepage}

\begin{center}

\vspace{1cm}

{\Large {\bf Sommerfeld Enhancements for Asymmetric Dark Matter }}

\vspace{1cm}

{\bf Aerman Sulitan, Hoernisa Iminniyaz, Mu Baoxia}\footnote{Corresponding 
author, wrns@xju.edu.cn}

\vskip 0.15in
{\it
{School of Physics Science and Technology, Xinjiang University, \\
Urumqi 830046, China} \\
}

\abstract{ We extend the analysis of asymmetric Dark Matter relic density with
  Sommerfeld enhancement to the case where the mediator 
  is massive. In asymmetric Dark Matter models, the asymmetric Dark Matter is 
  assumed to couple to light scalar or vector boson. Asymmetric Dark Matter
  annihilation cross section is enhanced by the Sommerfeld effect which exists
  due to the distortion of wavefunction of asymmetric Dark Matter
  particle and anti--particle by long--range interactions. The impacts of 
  Sommerfeld enhancement on the relic densities of
  asymmetric Dark Matter particle and anti--particle are discussed. The effect
  of kinetic decoupling on the relic density is also probed when the
  annihilation cross section is enhanced by Sommerfeld enhancement. Finally, 
  the constraints on the parameter space is given by using the observational 
  data of Dark Matter.}
\end{center}
\end{titlepage}
\setcounter{footnote}{0}

\section{Introduction}
 
The astrophysical and cosmological observations show that most of the 
matter in the universe is dark. The nature of Dark Matter (DM) is
not known to us although we have the precise value for DM relic density from 
the observations \cite{Ade:2015xua}. One assumption is that DM 
maybe asymmetric \cite{adm-models,frandsen,Petraki:2014uza}. 
The idea of asymmetric DM arises from the hypothesis that the 
present day abundance of DM may have the same origin with the visible 
or ordinary matter. The motivations come from the fact that the present day 
DM density is about 5 times of the average density of baryons 
$\Omega_{\rm DM} \simeq 5 \Omega_b$. 

In asymmetric DM models, it is often assumed that DM 
couples to light or massless force carriers 
\cite{Petraki:2014uza,Feng:2009mn,Agrawal:2017rvu,Agrawal:2016quu,Cirelli:2016rnw}. If
the mediator is light enough, the interaction between the asymmetric DM 
particle and anti--particle is
appeared as long-range. The wavefunctions of asymmetric
DM particles and anti--particles are distorted by the long--range
interaction if asymmetric DM interacts via the exchange of light mediators. 
It is the well-known Sommerfeld effect which enhances the
annihilation cross section of asymmetric DM particle and anti--particle 
\cite{Sommerfeld}. The Sommerfeld enhancement has effect on the relic 
density of asymmetric DM at some level. 

In ref.\cite{Abudurusuli:2020exx}, the authors explored the 
effect of Sommerfeld enhancement on the relic densities of asymmetric DM 
particle and anti--particle for $m_{\phi} = 0$. In 
ref.\cite{Abudurusuli:2020exx}, the Sommerfeld
enhancement factor $S_0$ is approximated by its value at $m_{\phi} = 0$, 
$S_0 = \pi \alpha /v/(1 - e^{-\pi\alpha/v})$, where $v$ is the velocity of 
asymmetric DM particle and anti--particle in the 
center--of--mass frame, $\alpha$ is the coupling strength. They obtained the 
result that 
the anti--particle abundance is significantly affected by the Sommerfeld 
effect comparing to the particle abundance. The decrease of abundance depends 
on the coupling strength $\alpha$. 
The impact of kinetic decoupling on the relic abundance of asymmetric DM is 
also analyzed when the annihilation cross section is changed by the Sommerfeld
enhancement in ref.\cite{Abudurusuli:2020exx}.

In refs.\cite{Feng:2010zp,Chen:2013bi}, the relic density
of symmetric DM is discussed when the light mediator mass $m_{\phi} \neq 0$,
where $m_{\phi} \ll m$ with $m$ being the DM mass. 
In our work, we extend this exploration to the asymmetric DM case. For massive
mediator, the Sommerfeld enhancement is saturated at low velocity, and exhibits
resonant behavior at some specific values of 
$m_{\phi}$ \cite{ArkaniHamed:2008qn}. Sommerfeld enhancement boosts the 
late--time DM annihilation signals
\cite{Pospelov:2008jd,MarchRussell:2008tu}. The coupling of asymmetric DM to 
the light force mediator determines the Sommerfeld enhancement . 
The coupling for asymmetric DM needs to be stronger than the 
symmetric DM of the same mass. Therefore, the importance of 
Sommerfeld enhancement for the phenomenology of asymmetric DM may be 
more significant than the symmetric DM case. 
We investigate the quantitative impact of Sommerfeld enhancement on the 
asymmetric DM relic density for the case of $m_\phi \neq 0 $. When the 
asymmetric DM particles and anti--particles decoupled from the chemical 
equilibrium, they were still in kinetic equilibrium for a while due to the 
scattering of standard 
model particles. The background radiation temperature and the temperatures of 
asymmetric DM particle and anti--particle are different before and after 
kinetic decoupling \cite{Bringmann:2006mu,Bringmann:2009vf}. This difference 
leads to significant change of the relic
density of asymmetric DM after kinetic decoupling. It results that the relic 
density of asymmetric DM is continuously decreased until the Sommerfeld 
enhancement is saturated at low velocity. The decrease is more sizable for
asymmetric DM anti--particle than the particle. In our work, we only consider 
the Sommerfeld enhancement and 
neglect the effect of bound state formation on the relic density of asymmetric 
DM. The bound--state formation affects the relic density of DM only around the
unitarity bound \cite{Cirelli:2016rnw,vonHarling:2014kha}.  

The paper is arranged as following. In section 2, we discuss the effect of 
Sommerfeld enhancement on the asymmetric DM abundance. In section 3, we use
the Planck data to obtain the constraints on parameter spaces. 
The last section is devoted to the conclusions.

\section{Effect of Sommerfeld enhancement on the asymmetric DM abundance }
Asymmetric DM annihilation process may be due to the long-range interactions 
mediated by light mediator. If asymmetric DM couples to light force mediator,
the wavefunction of asymmetric DM particle and anti--particle is distorted
by the long--range interaction. It is the Sommerfeld effect \cite{Sommerfeld}.  
This results the annihilation cross section of asymmetric DM is enhanced at
low velocity. The Sommerfeld enhanced annihilation cross section affects the
relic density of asymmetric DM. 

When the mediator mass $m_{\phi} \neq 0$, the 
analytic approximation of the Sommerfeld enhancement factor for $s-$wave 
annihilation \cite{Cassel:2009wt,Slatyer:2009vg,Feng:2010zp,Chen:2013bi} is  
\begin{equation}\label{eq:Som_s}
S = \frac{\pi }{\epsilon_v} 
   \frac{{\rm sinh}\left(\frac{2 \pi \epsilon_v}{\pi^2 \epsilon_\phi/6 }\right)}
   {{\rm cosh}\left(\frac{2 \pi \epsilon_v}{\pi^2 \epsilon_\phi/6 }\right) - 
    {\rm cos}\left(2\pi\sqrt{\frac{1}{\pi^2 \epsilon_\phi/6} - 
      \frac{ \epsilon^2_v}{(\pi^2 \epsilon_\phi/6)^2}}\right)}\,,  
\end{equation}
where $\epsilon_v \equiv v/\alpha$ and 
$\epsilon_\phi \equiv m_\phi /(\alpha m)$. For $\epsilon_\phi \gg \epsilon_v$, 
there are resonances at
\begin{equation} \label{eq:reson}
      m_{\phi} \simeq \frac{6 \alpha m}{\pi^2 n^2},
\end{equation}
here $n$ is the positive integer. The Sommerfeld enhancement factor for 
low $v$ is $S \simeq \pi^2 \alpha m_\phi/(6 m v^2)$ at theses resonances with 
$m_{\phi}$ given by Eq.(\ref{eq:reson}).

Following, we discuss the impact of Sommerfeld enhancement on the relic 
densities of asymmetric DM particle and anti--particle in
model independent way. Usually it is assumed the asymmetric DM particles and
anti--particles were in thermal equilibrium with the standard model particles
in the radiation dominated universe. When the annihilation rate 
$\Gamma_{\rm an} = n_{\chi,\bar\chi} \langle \sigma v_{\rm rel}\rangle$ is
less than the expansion rate $H$, the particles and anti--particles can not 
keep in the chemical equilibrium and decouple from thermal plasma. Here, 
 $n_{\chi,\bar\chi}$ are the number densities of asymmetric DM
particle $\chi$ and anti--particle $\bar\chi$, 
$\langle \sigma v_{\rm rel} \rangle $ is the
thermal average of the annihilation cross section $\sigma$ multiplied with the
relative velocity $v_{\rm rel}$ of asymmetric DM particle and anti--particle, 
$v = v_{\rm rel}/2$.
The temperature at which the decoupling occurred is called freeze out 
temperature. The asymmetric DM
abundance is almost fixed at the freeze out temperature 
\cite{GSV,Iminniyaz:2011yp,standard-cos}. To 
determine the relic density $\Omega_{\rm DM}$ of DM , we 
need to solve the Boltzmann equations which describe the particle and
anti--particle evolution in the universe.  

The Boltzmann equations expressed by the ratio of $n_{\chi,\bar\chi}$ to
entropy density $s$, 
$Y_{\chi,\bar\chi} =n_{\chi,\bar\chi}/s$ with $s= 2 \pi^2 g_{*s}/45\, T^3$, are
\begin{eqnarray} \label{eq:boltzmann_Y}
\frac{d Y_{\chi}}{dx}
    = - \frac{\langle \sigma v_{\rm rel} \rangle s }{H x}\,
     (Y_{\chi}~Y_{\bar\chi} - Y_{\chi, {\rm eq}}~Y_{\bar\chi, {\rm eq}} )\,,
\end{eqnarray}
\begin{eqnarray} \label{eq:boltzmann_Ybar}
\frac{d Y_{\bar{\chi}}}{dx}
    = - \frac{\langle \sigma v_{\rm rel} \rangle s }{H x}\,
     (Y_{\chi}~Y_{\bar\chi} - Y_{\chi, {\rm eq}}~Y_{\bar\chi, {\rm eq}} )\,,
\end{eqnarray}
where $x=m/T$, and the expansion rate is $H = \pi T^2 \sqrt{g_*/90}/M_{\rm Pl}$,
here $g_{*s}$, $g_*$ are the effective number of entropic degrees of
freedom and relativistic degrees of freedom. $M_{\rm Pl} = 2.4\times
10^{-18}$GeV is the reduced Planck mass. Here we assume only the asymmetric 
Dark Matter particle $\chi$ and anti--particle $\bar\chi$ 
annihilate into standard model particles. Thermal average of the Sommerfeld 
enhanced annihilation cross section is 
\begin{equation}
   \langle \sigma v_{\rm rel}\rangle = \sigma_0 \langle S(v_{\rm rel})\rangle\,,  
\end{equation}
here $\sigma_0$ corresponds to the s--wave annihilation cross 
section, 
\begin{equation}
    \langle S(v_{\rm rel})\rangle = 
      \frac{x^{3/2}}{2 \sqrt{\pi}}\,\int^{\infty}_0 \,
          v_{\rm rel}^2 ~ e^{-\frac{x}{4} v_{\rm rel}^2}\, S\,dv_{\rm rel} \, .
\end{equation}
The equilibrium abundances for the asymmetric DM particle and
anti--particle are 
\begin{equation}\label{eq:boltzmann_eq}
Y_{\chi,{\rm eq}} = \frac{90}{(2\pi)^{7/2}}\frac{g_{\chi}}{g_{*s}}\, 
                x^{3/2} {\rm e}^{-x(1-\mu_{\chi}/m)}\,, 
\end{equation}
\begin{equation}\label{eq:boltzmann_eqbar}
Y_{\bar\chi,{\rm eq}} = \frac{90}{(2\pi)^{7/2}}\frac{g_{\chi}}{g_{*s}} \,
                   x^{3/2} {\rm e}^{-x(1+\mu_{\chi}/m)}\,,
\end{equation}
where we used the fact that chemical potential $\mu_{\bar\chi} = - \mu_{\chi}$ in 
equilibrium state, and $g_{\chi}$ is the number of intrinsic degrees of 
freedom of the particle. Subtracting 
Eq.(\ref{eq:boltzmann_Ybar}) from 
Eq.(\ref{eq:boltzmann_Y}), we obtain 
$ d Y_{\chi}/dx - d Y_{\bar\chi}/dx =0$. This requires 
$ Y_{\chi} - Y_{\bar\chi} = \eta$, where $\eta$ is a constant.
Then the Boltzmann equations are rewritten as
\begin{equation} \label{eq:Yeta}
\frac{d Y_{\chi}}{dx} =
     - \frac{\lambda \langle S(v_{\rm rel}) \rangle}{x^2}~  
     (Y_{\chi}^2 - \eta Y_{\chi}  - Y^2_{\rm eq})\,,
\end{equation}
\begin{equation} \label{eq:Ybareta}
\frac{d Y_{\bar\chi}}{dx} =
     - \frac{\lambda \langle S(v_{\rm rel}) \rangle}{x^2}~  
     (Y_{\bar\chi}^2 + \eta Y_{\bar\chi}  - Y^2_{\rm eq})\,,
\end{equation}
here 
$Y^2_{\rm eq}= Y_{\chi,{\rm eq}} Y_{\bar\chi,{\rm eq}}=(0.145g_{\chi}/g_*)^2\,x^3e^{-2x}$ 
and $\lambda = 1.32\,m M_{\rm Pl}\,\sigma_0 \, \sqrt{g_*}$, where we assume 
$g_*\simeq g_{*s}$ and $dg_*/dx\simeq 0$.

Eqs.(\ref{eq:Yeta}) and (\ref{eq:Ybareta}) can be solved numerically. We can
also obtain the semi--analytical solution by repeating the same method which 
is used in ref.\cite{Iminniyaz:2011yp}. First, we can have the semi--analytic 
result of Eq.(\ref{eq:Ybareta}). We use the deviation 
$\Delta_{\bar\chi} = Y_{\bar\chi} - Y_{\bar\chi, {\rm eq}}$ to express the
Boltzmann equation (\ref{eq:Ybareta}) as  
\begin{equation} \label{eq:deltabar}
      \Delta_{\bar\chi}^{\prime} = - Y^{\prime}_{\bar\chi, {\rm eq}} - 
        \lambda x^{-2} \langle S(v_{\rm rel}) \rangle\, 
       \Delta_{\bar\chi}( 2 Y_{\bar\chi,{\rm eq}} + \Delta_{\bar\chi}
      + \eta )\,,
\end{equation}
here $\prime$ denotes $d/dx$ and 
$Y_{\bar\chi,{\rm eq}} = - \eta/2 + \sqrt{\eta^2/4 + Y^2_{\rm eq}}$, which is
obtained by using the fact that right hand side of equation 
(\ref{eq:boltzmann_Ybar}) vanishes in equilibrium. When the temperature is 
high, the value of $Y_{\bar\chi}$ tracks its equilibrium value very closely, 
and $\Delta^2$ and $\Delta^{\prime}$ are negligible. Then we obtain
\begin{equation} \label{eq:bardelta_solu}
      \Delta_{\bar\chi} \simeq \frac{2 x^2 \, Y^2_{\rm eq}}{   
     \lambda \langle S(v_{\rm rel}) \rangle\,(\eta^2 + 4 Y^2_{\rm eq})}\,.
 \end{equation}
The Sommerfeld effect is not significant up to the time of freeze out, 
therefore we use standard method to fix the freeze out temperature.  
The freeze out temerature is defined as following,
\begin{equation}
      \Delta_{\bar\chi, {\rm ns}}(\bar{x}_f) = \xi\, Y_{\bar\chi,{\rm eq}}(\bar{x}_f)\,,
\end{equation}
where $\Delta_{\bar\chi, {\rm ns}}(\bar{x}_f) 
 \simeq 2 \bar{x}_f^2 \, Y^2_{\rm eq}/ [\lambda (\eta^2 + 4 Y^2_{\rm eq})]$,
which is indeed Eq.(\ref{eq:bardelta_solu}) without including Sommerfeld 
factor, here $\bar{x}_f$ is the freeze out temperature for asymmetric DM
anti--particle, it results 
\begin{equation} \label{eq:xf}
      2 \bar{x}_f^2 \, Y^2_{\rm eq}/   
     [\lambda \, (\eta^2 + 4 Y^2_{\rm eq})]
      = \xi\, Y_{\bar\chi,{\rm eq}}(\bar{x}_f)\,,
\end{equation}
and $\xi = \sqrt{2} -1$ \cite{standard-cos}. $\bar{x}_f$ is determined by 
iteratively solving Eq.(\ref{eq:xf}). 

For low temperature, $Y_{\bar\chi,{\rm eq}}$ is negligible. Dropping the terms
related to $Y_{\bar\chi,{\rm eq}}$, then Eq.(\ref{eq:deltabar}) becomes
\begin{equation} \label{eq:deltalate}
 \Delta^{\prime}_{\bar\chi} = - \lambda x^{-2} \langle S(v_{\rm rel}) \rangle\, 
       \Delta_{\bar\chi}( \Delta_{\bar\chi} + \eta )   \,.
\end{equation}

Although the asymmetric
DM particles and anti--particles decoupled from the chemical equilibrium, they
were still in kinetic equilibrium for a while due to the scattering of the
standard model particles. When the particles and anti--particles were in
chemical and kinetic equilibrium, the temperatures $T_{\chi,\bar\chi}$ of 
asymmetric DM track the background radiation temperature $T$, 
$T_{\chi,\bar\chi} = T$. After kinetic
decoupling, the temperatures of asymmetric DM particle and anti--particle 
scale as $T_{\chi,\bar\chi} = T^2/T_k $, where $T_k$ is the kinetic decoupling
temperature \cite{Bringmann:2006mu,Bringmann:2009vf,Chen:2001jz,Hofmann:2001bi}. The change of temperature 
of asymmetric DM before and after kinetic decoupling leaves its impact on the 
relic density of asymmetric DM. With the kinetic decoupling, the thermal 
average of Sommerfeld enhancement factor takes the form as
\begin{equation}
    \langle S_k(v_{\rm rel})\rangle = 
      \frac{x^3}{2 \sqrt{\pi}\,x^{3/2}_k}\,\int^{\infty}_0 \,
          v_{\rm rel}^2 ~ e^{-\frac{x^2}{4x_k} v_{\rm rel}^2}\, S\,dv_{\rm rel} ,
\end{equation}
where $x_k = m/T_k$. 

Integrating Eq.(\ref{eq:deltalate}), we obtain the relic 
abundance of asymmetric DM anti--particle.
The integration range should be separated into two 
parts after including kinetic decoupling: the first part is for the era before 
kinetic decoupling, the second part is for the era of after kinetic decoupling,
then 
\begin{equation} \label{eq:barY_cross1}
Y_{\bar\chi}(x_s) =  \eta\,\left\{ \,
  {\rm exp} \left[\, 1.32\, \eta \, m M_{\rm Pl} \sigma_0\,
     \sqrt{g_*} \,\left( \int^{x_k}_{\bar{x}_f} 
     \frac{\langle S(v_{\rm rel}) \rangle}
          { x^2}dx + \int^{x_s}_{x_k} 
     \frac{\langle S_{\rm k}(v_{\rm rel}) \rangle}
          { x^2}dx\, \right) \right] -1\, \right\}^{-1} \,,
\end{equation}
here $x_s$ is the point at which
the Sommerfeld enhancement saturates at low velocity. We assume that 
$\Delta_{\bar\chi}(\bar{x}_f)\gg \Delta_{\bar\chi}(x_s)$.

The asymmetric DM particle abundance is obtained by using the relation 
$ Y_{\chi} - Y_{\bar\chi} = \eta $,
\begin{equation} \label{eq:barY_cross2}
Y_{\chi}(x_s) =  \eta\,\left\{ \, 1 -
  {\rm exp} \left[\,- 1.32\, \eta \, m M_{\rm Pl} \sigma_0\,
     \sqrt{g_*} \,\left( \int^{x_k}_{x_f} 
     \frac{\langle S(v_{\rm rel}) \rangle}
          { x^2}dx + \int^{x_s}_{x_k} 
     \frac{\langle S_k(v_{\rm rel}) \rangle}
          { x^2}dx\, \right) \right] \, \right\}^{-1} \,,
\end{equation}
where $x_f$ is the inverse scaled freeze out temperature for $\chi$. 
Eqs.(\ref{eq:barY_cross1}) and (\ref{eq:barY_cross2}) are only consistent with
the constraint $ Y_{\chi} - Y_{\bar\chi} = \eta $ if $x_f = {\bar x}_f$. 

The final DM relic density is given by 
\begin{equation}
      \Omega_{\rm DM} h^2 \simeq 2.76 \times 10^8\,m 
              \left[ Y_{\chi}(x_s) + Y_{\bar\chi}(x_s) \right]\,,
\end{equation}
where $ h = 0.673 \pm 0.098 $ is the scaled Hubble constant in units of 
$100$ km s$^{-1}$ Mpc$^{-1}$ and $\Omega_{\chi} = \rho_{\chi}/\rho_c$. Here 
$\rho_{\chi}=n_{\chi} m = s_0 Y_{\chi}  m$ and the
critic density is $\rho_c = 3 H^2_0 M^2_{\rm Pl}$, 
where $s_0 \simeq 2900$ cm$^{-3}$ is the present entropy density and $H_0$ is
the Hubble constant.
%

\begin{figure}[h]
  \begin{center}
     \hspace*{-0.5cm} \includegraphics*[width=8cm]{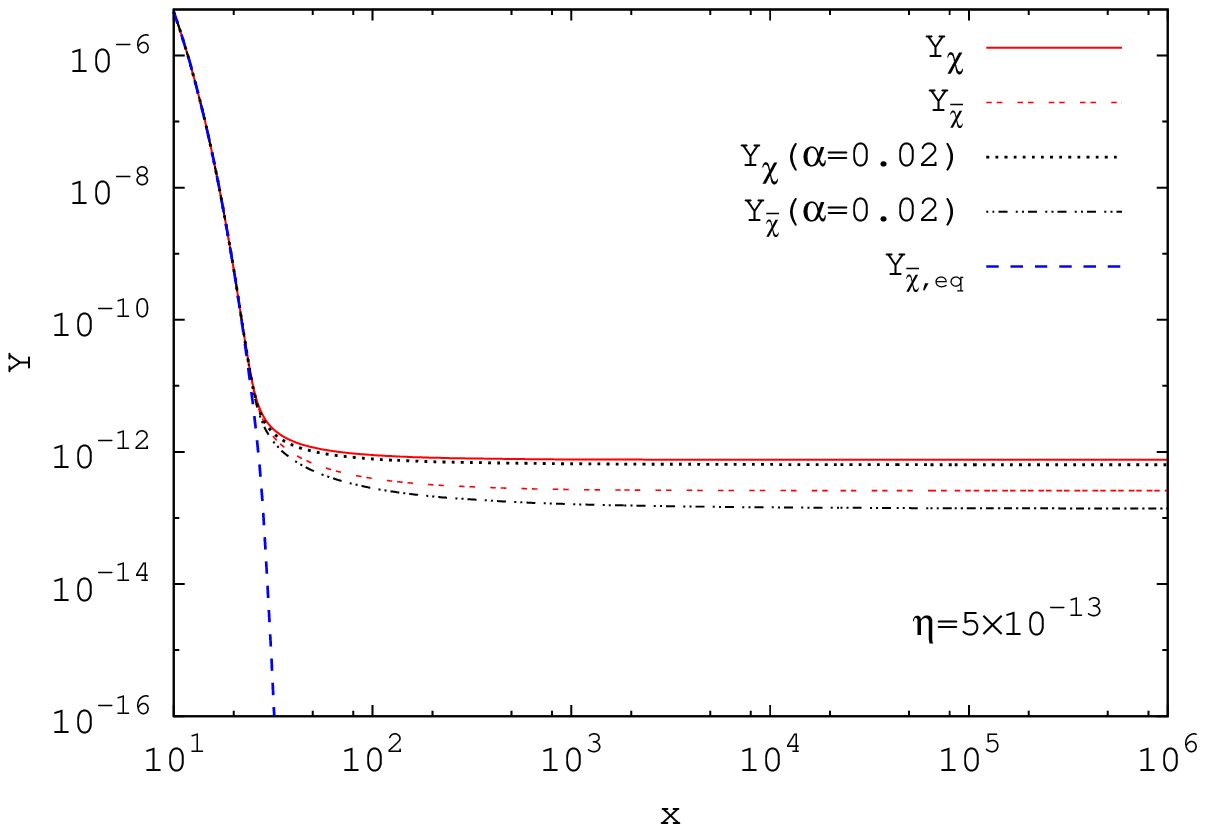}
    \put(-115,-12){(a)}
    \hspace*{-0.5cm} \includegraphics*[width=8cm]{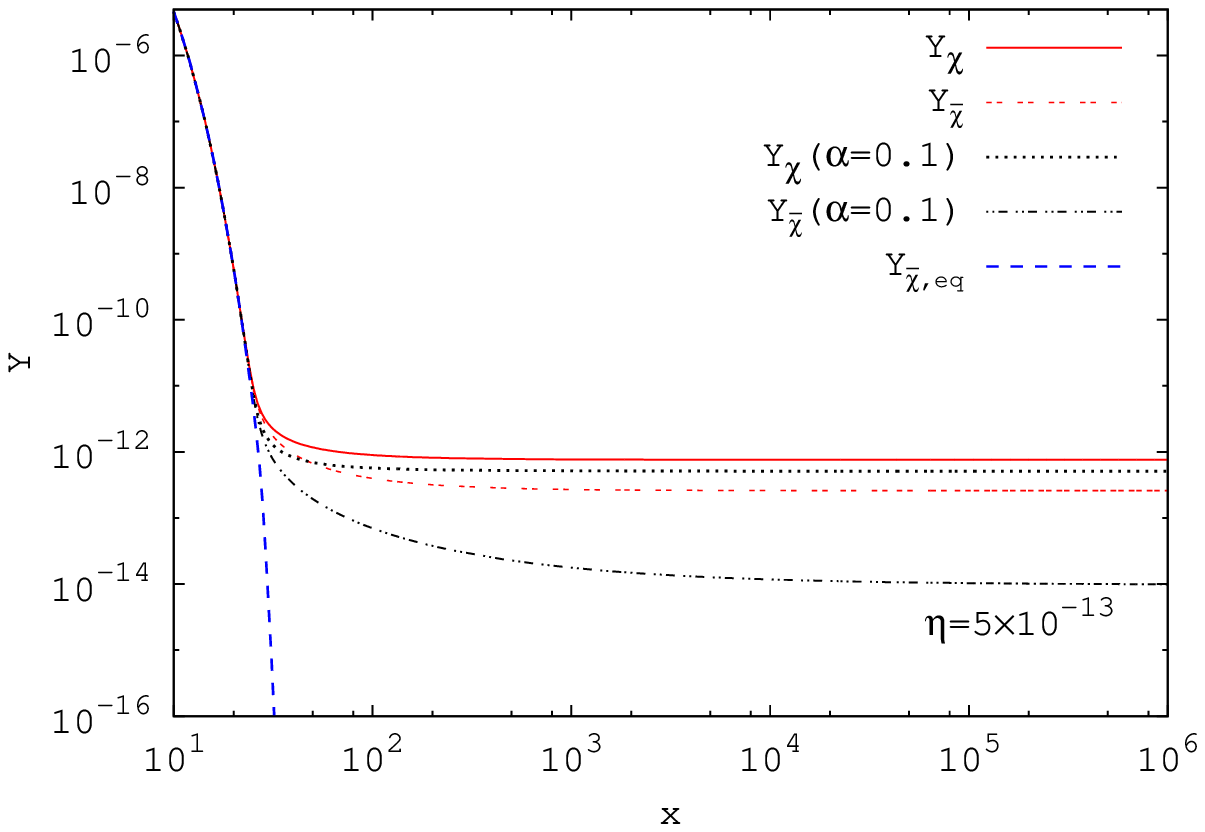}
    \put(-115,-12){(b)}
     \vspace{0.5cm}
     \hspace*{-0.5cm} \includegraphics*[width=8cm]{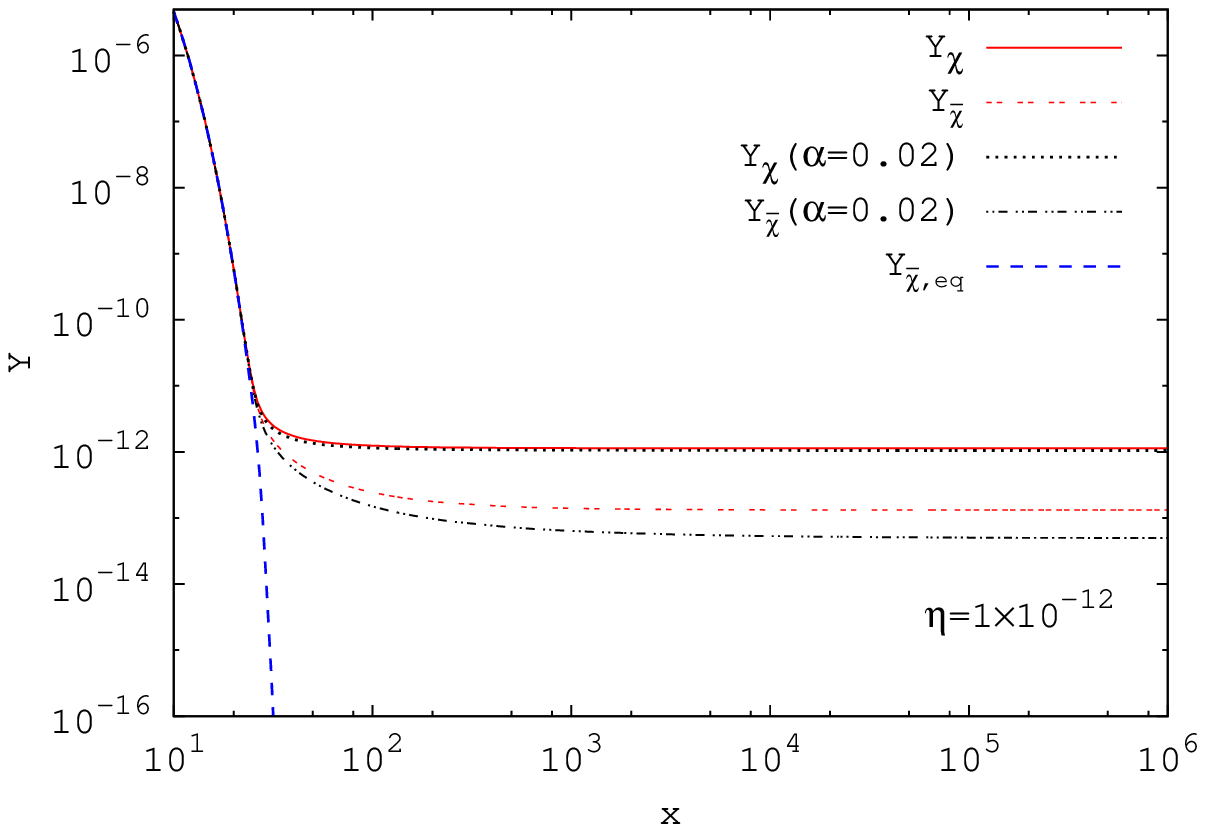}
    \put(-115,-12){(c)}
    \hspace*{-0.5cm} \includegraphics*[width=8cm]{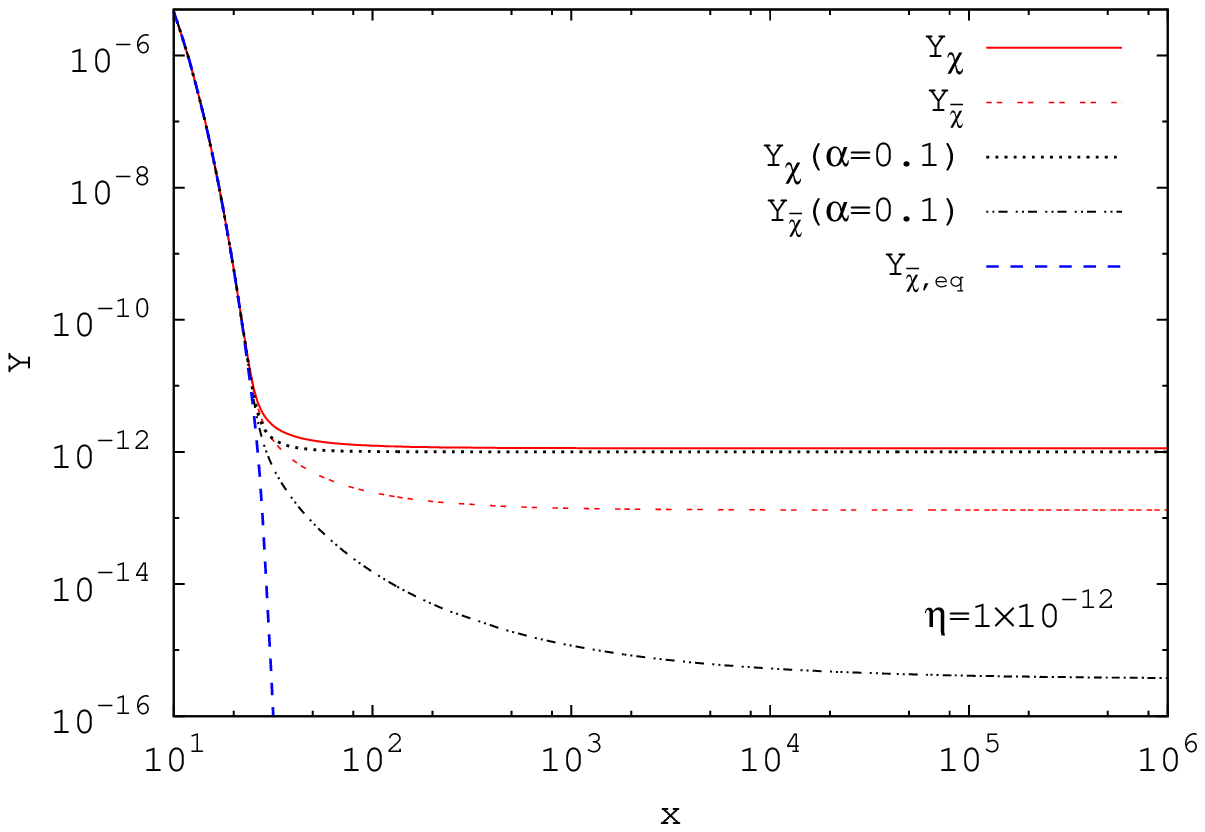}
    \put(-115,-12){(d)}
     \caption{\label{fig:a} \footnotesize
  Asymmetric DM abundances $Y_{\chi}$ and $Y_{\bar\chi}$ as a function of $x$
  for the Sommerfeld enhanced annihilation cross section. Here 
   $\sigma_0 = 4 \times 10^{-26}$ ${\rm cm}^3 \,{\rm s}^{-1}$, $m = 500$ GeV, 
   $x_f = 25 $, $m_{\phi} = 0.25$ GeV, $g_{\chi} = 2$, $g_* = 90$.}  
      \end{center}
\end{figure}
We use the numerical solutions of equations (\ref{eq:Yeta}), 
(\ref{eq:Ybareta}) to plot the evolution of 
asymmetric DM abundance as a function of the inverse--scaled 
temperature $x$ for $\alpha = 0.02$ and $\alpha = 0.1$ in 
Fig.\ref{fig:a}. Here $\eta = 5 \times 10^{-13}$ in panels (a) and (b), 
$\eta = 1 \times 10^{-12}$ in (c) and (d). The two thick and dashed (red)
lines are for the asymmetric DM particle and anti--particle abundances 
without Sommerfeld enhancement; the dotted and dash--dotted (black) lines 
are the abundances of $Y_{\chi}$ and $Y_{\bar\chi}$ with Sommerfeld
enhancement when the coupling strength $\alpha = 0.02$ in (a), (c), 
$\alpha = 0.1$ in (b), (d); the dashed (blue) line is for the equilibrium
value of anti--particle abundance. Here $m_{\phi} = 0.25$ GeV. The annihilation
cross section is enhanced because of the Sommerfeld effect. The abundances of 
asymmetric DM particle and anti--particle are decreased due to the 
enhanced annihilation cross section. The size 
of the decrease depends on coupling strength $\alpha$. For example, the 
decreases of abundances $Y_{\chi}$ and $Y_{\bar\chi}$ are larger for 
$\alpha = 0.1$ in panel $(b)$, $(d)$ than the case of
$\alpha = 0.02$ in $(a)$, $(c)$ in Fig.\ref{fig:a}. 
The reduction of anti--particle abundance is siginificant for larger coupling
strength comparing to the case of particle abundance.
When the asymmtry factor is smaller as $5 \times 10^{-13}$, the difference of 
asymmetric DM particle abundance with and without Sommerfeld enhancement is 
visible which is shown in panle $(b)$ of Fig.\ref{fig:a}.
For larger asymmetry factor as $1 \times 10^{-12}$, the decrease of particle 
abundance is not visible.

\begin{figure}[h]
  \begin{center}
    \hspace*{-0.5cm} \includegraphics*[width=8cm]{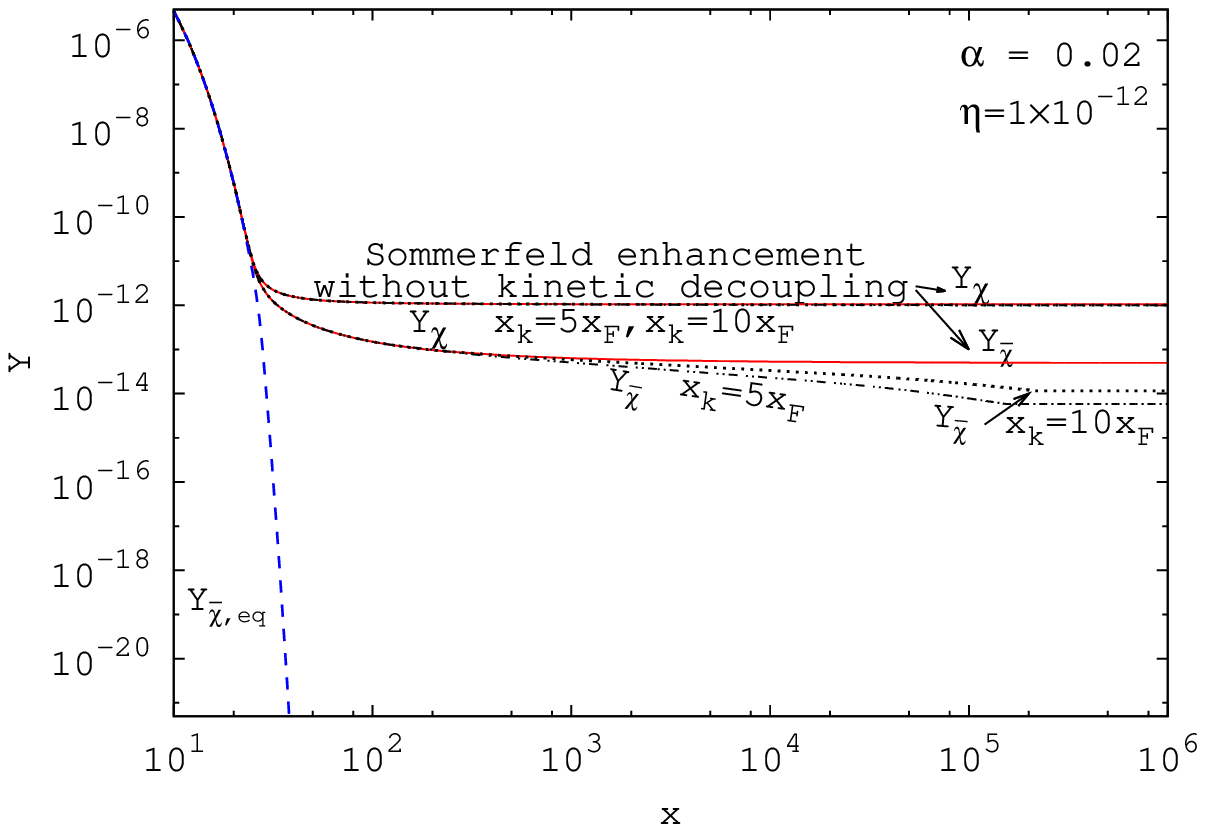}
    \put(-115,-12){(a)}
    \hspace*{-0.5cm} \includegraphics*[width=8cm]{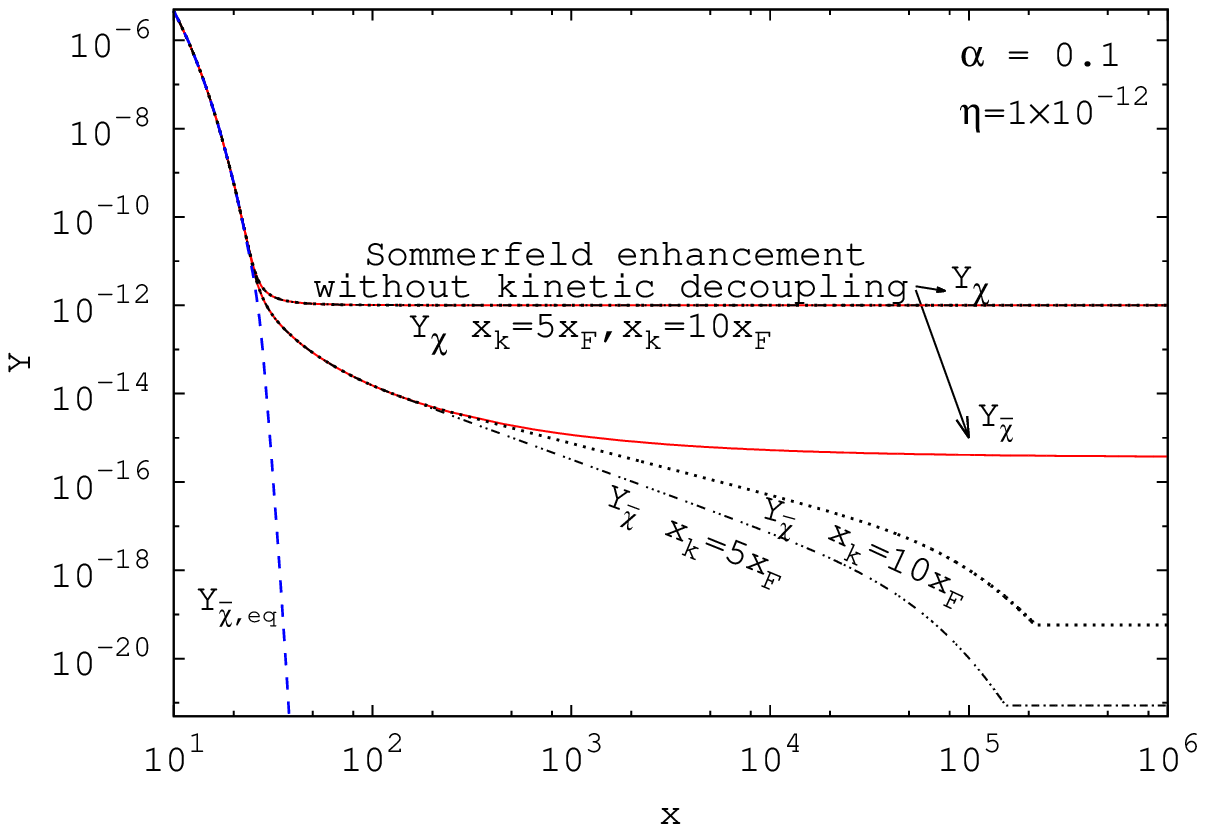}
    \put(-115,-12){(b)}
     \caption{\label{fig:b} \footnotesize Evolution of $Y$ for the particle
       and anti--particle as a function of $x$ for the case when there is 
       Sommerfeld enhancement with kinetic decoupling. 
       Here $\sigma_0 = 4 \times 10^{-26}$ ${\rm cm}^3 \,{\rm s}^{-1}$,
       $m = 500$ GeV, $x_f = 25 $, 
    $m_{\phi} = 0.25$ GeV, $g_{\chi} = 2$, $g_* = 90$. }
  \end{center}
\end{figure}
\begin{figure}[h!]
  \begin{center}
    \hspace*{-0.5cm} \includegraphics*[width=8cm]{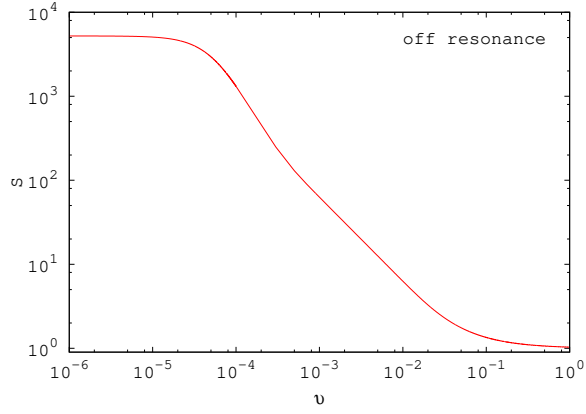}
     \caption{\label{fig:bb} 
     \footnotesize Sommerfeld enhancement factor $S$ as a function of the
     velocity $v$ for $\alpha = 0.02$, $m_{\phi} = 0.25$ GeV, $m = 500$ GeV.  }
  \end{center}
\end{figure}

The relic abundances of asymmetric DM with Sommerfeld enhancement including the
effect of kinetic decoupling is plotted in Fig.\ref{fig:b} for $\alpha = 0.02$
and $\alpha = 0.1$. These plots are based on the numerical solutions of equation
(\ref{eq:Yeta}), (\ref{eq:Ybareta}) with the seperated integration range. The
integration ranges are from $x_f$ to $x_k$ when there is only Sommerfeld
enhancement and from $x_k$ to $x_s$ while there is kinetic
decoupling. Here $\eta = 1 \times 10^{-12}$. The dotted (black) line
is for anti--particle DM abundance for $x_k = 10 x_f$ and dashed (black) 
line is for $x_k= 5 x_f$. We found the relic abundance of asymmetric DM 
anti--particle is continuously
decreased after kinetic decoupling. When the kinetic decoupling 
temperature is closer to the freeze out temperature, the decrease is larger.
 In panel (a), around $x = 1.5 \times 10^5$ for $x_k= 5x_f$ and 
$x = 2.5 \times 10^5$ for $x_k= 10x_f$, the curves become completely flat. It
means $Y_{\bar\chi}$ is constant. At this point, Sommerfeld enhancement
is saturated. We can see the reason from Fig.\ref{fig:bb}. When the velocity
is small, the Sommerfeld factor $S$ goes to contant value. This is the reason 
why $Y_{\bar\chi}$ becomes constant. Here we take 
$\alpha = 0.02$, $m_{\phi} = 0.25$ GeV, $m = 500$ GeV in Fig.\ref{fig:bb}. We
took the parameter set in the paper where the Sommerfeld enhancement is 
not near a resonance. In panel (b), 
anti--particle abundance becomes stable from $x = 1.6 \times 10^5$ for 
$x_k= 5x_f$ and $x = 2.2 \times 10^5$ for $x_k= 10x_f$.
The decrease of asymmetric DM particle abundances are almost invisible in
these plots. Asymmetric DM particle abundances for the two cases are
overlapped with the case of without kinetic decoupling.

\section{Constraints}
We have the  Planck data which provides the Dark Matter relic density as 
\cite{Ade:2015xua},
\begin{eqnarray} \label{eq:pldata}
  \Omega_{\rm DM} h^2 = 0.1199 \pm 0.0022\, .
\end{eqnarray}
\begin{figure}[h]
  \begin{center}
    \hspace*{-0.5cm} \includegraphics*[width=8.7cm]{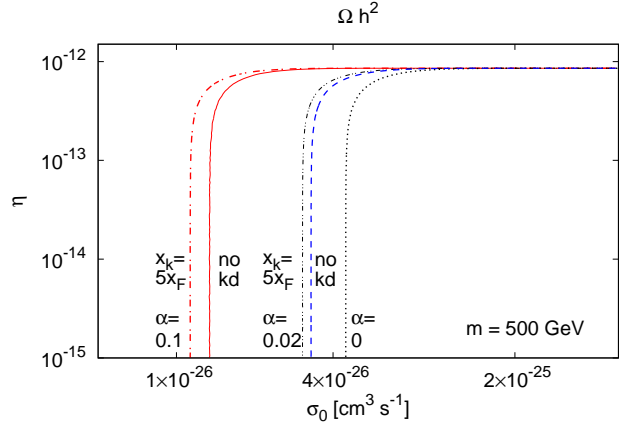}
    \caption{\label{fig:c} 
    \footnotesize Contour plot of $s-$wave annihilation cross section 
    $\sigma_0$ and asymmetry factor $\eta$ when 
    $\Omega_{\rm DM} h^2 = 0.1199$. Here $m = 500$ GeV, $x_f = 25 $, 
    $m_{\phi} = 0.25$ GeV,
    $g_{\chi} = 2$, $g_* = 90$.}
     \end{center}
\end{figure}
The contour plot of $s-$wave annihilation cross section $\sigma_0$ and 
asymmetry factor $\eta$ is shown in Fig.\ref{fig:c} when 
$\Omega_{\rm DM} h^2 = 0.1199$. The dotted (black) line is for the case
of no Sommerfeld enhancement; long--dashed (blue) and double dot--dashed 
(black) lines correspond to the case of Sommerfeld enhancement without 
and with kinetic decoupling for $\alpha = 0.02$; 
thick (red) and dot--dashed (red) lines are for $\alpha = 0.1$, here 
$x_k = 5 x_F$. We found the
required annihilation cross section with the Sommerfeld enhancement is smaller
than the case of without. There is less relic density due to the enhanced
cross section. Therefore, the needed cross section should be smaller in 
order to satisfy the observed range of DM relic density. When there is kinetic
decoupling, the required annihilation cross section is smaller than the case
of no kinetic decoupling. The relic density of asymmetric DM is continuously 
decreased after the kinetic decoupling until the Sommerfeld enhancement
saturates at small velocities. It results the required cross section for 
kinetic decoupling is smaller than the case of no kinetic decoupling when the
asymmetry factor is small. 
For larger asymmetry factor, the required cross sections with kinetic 
decoupling and without are the same. The relic density is only 
determined by the asymmetry factor in that case.

\section{Summary and conclusions}

The effect of Sommerfeld enhancement on the relic density of asymmetric DM 
is discussed in this work. Here we generalize the case of massless force 
mediator to the massive case. The cross section between the asymmetric DM 
particle and anti--particle is enhanced by the Sommerfeld effect. 
We investigate in which extent the relic densities of
asymmetric DM particle and anti--particle are affected when the asymmetric DM
annihilation cross section is enhanced by the Sommerfeld effect. We found the
asymmetric DM particle and anti--particle abundances are decreased due to the
enhancement of annihilation cross section. The reduction
of anti--particle relic abundance is more significant than particle
abundance. The decrease depends on the size of coupling strength $\alpha$. For 
larger coupling strength, the decrease is larger.    

After asymmetric DM particles and anti--particles decoupled from the chemical
equilibrium, they were still in kinetic equilibrium for a while because of 
the scattering off relativistic standard 
model particles in the thermal plasma. The asymmetric DM particles and
anti--particles decouple from the kinetic equilibrium when the scattering rate 
falls below the expansion rate. In our work, we explore the effects of kinetic
decoupling on the relic density when there is Sommerfeld enhancement. We found
the relic abundances of asymmetric DM are decreased continuously until the 
Sommerfeld enhancement saturates. The reduction of  
anti--particle abundance is significant than the particle abundance. The level 
of decrease depends on the kinetic decoupling temperature, coupling 
strength $\alpha$. For example, there is larger decrease when the kinetic 
decoupling temperature is more close to the freeze out point.

Finally, using the observed DM abundance,  we obtain the constraints on the 
annihilation cross section and asymmetry factor when there is Sommerfeld 
enhancement. We found the required annihilation cross section with 
Sommerfeld enhancement is smaller than the case of without. Also the wanted 
annihilation cross section is smaller for the case of kinetic decoupling than 
the case of no kinetic decoupling. Those results are significant for
asymmetric DM when Sommerfeld effect is important at low 
velocity. Sommerfeld effect implies notable indirect detection signals
from asymmetric DM anti--particle. Therefore, we have the possibility to 
explore the asymmetric DM by the observation of CMB (Cosmic Microwave 
Background), the Milky way and Dwarf galaxies.

\section*{Acknowledgments}

The work is supported by the National Natural Science Foundation of China
(11765021).

\end{document}